\documentclass[twocolumn,showpacs,   
               preprintnumbers,     
               aps,                 
               prd,          	    
               groupedaddress,      
               nofootinbib,         
               tightenlines,        
               floats,floatfix      
               ]{revtex4}

\usepackage{graphicx}
\usepackage{latexsym}
\usepackage{amsmath,amssymb}        
\usepackage[draft=false]{hyperref}

\begin{document}



\title{Evolving spherical Boson Stars on a 3D cartesian grid}
\author{F. Siddhartha Guzm\'an\footnote{E-mail: guzman@cct.lsu.edu}}
\affiliation{
Max Planck Institut f\"ur Gravitationsphysik, Albert Einstein 
Institut, Am M\"uhlenberg 1, 14476 Golm, Germany.\\
Center for Computation and Technology, 302 Johnston Hall, Louisiana State 
University, Baton Rouge, LA 70803.\footnote{Current address.}} 
\date{\today}
\pacs{04.40.-b, 04.25.Dm, 04.30.Db}


\begin{abstract}
A code to evolve boson stars in 3D is presented as the starting point
for the evolution of scalar field systems with arbitrary symmetries. It 
was possible to reproduce the known results related to perturbations 
discovered with 1D numerical codes in the past, which include evolution of 
stable and unstable equilibrium configurations. In addition, the apparent 
and event horizons masses of a collapsing boson star are shown for the 
first time. The present code is expected to be useful at evolving possible 
sources of gravitational waves related to scalar field objects and to 
handle toy models of systems perturbed with scalar fields in 3D.
\end{abstract}

\maketitle


\section{Introduction}

Boson Stars (BSs) are localized, regular, spherically symmetric solutions 
of Einstein's field equations, whose matter content is provided by a 
complex scalar field with mass and self-interaction \cite{ruffini,kaup,colpi}.
Recently Boson Stars have appeared in many contexts. Traditionally they 
were proposed to be alternatives of compact objects \cite{ruffini,colpi}; 
they have been proposed as candidates for dark matter \cite{dark-matter}; 
this includes the possibility to emulate galactic supermassive Black Holes 
\cite{diego}; they have even been considered as sources of gravitational 
waves \cite{ryan}; and they have been used in the field of numerical 
relativity to test evolution formulations of Einstein's equations and 
experience with gauge conditions \cite{towards}, since these objects are 
very tractable in the sense that they have no defined surface and their 
dynamics do not tend to form shocks, and therefore sophisticated 
techniques -like shock capturing methods- are not required. Moreover, BSs 
define sequences of equilibrium configurations which serve to test a 
numerical implementation. That is, for a spherical equilibrium 
configuration the field has to oscillate with a constant frequency while 
the geometry of the space-time has to remain static. For general reviews 
in BSs see \cite{jetzer,cqgreview}. In this paper I focus on numerical 
challenges, and set up a starting point to study the astrophysical 
applications mentioned before. Instead of developing the original mean 
field approximation to calculate the expectation value of the 
stress-energy tensor components of a quantum scalar field as done 
originally \cite{ruffini,kaup}, the field is simply considered to be 
classical as usually done when studying the evolution of these systems.\\

For several reasons, such systems have only been evolved for long times
(long enough to observe the oscillations of a perturbed star) in 
spherical symmetry using 1D codes \cite{seidel90,seidel98,scott2000}. The 
main reason for this is without doubt the difficulty in formulating the 
evolution equations for the space-time from Einstein's equations when 
using non-spherical symmetry. In spherical symmetry these evolution 
equations are replaced with the slicing condition and the constraints
(see for instance \cite{seidel98,scott2000}). The resulting ODEs can be 
integrated at each time step during the evolution, which is 
computationally cheap and allows a constrained evolution under control for 
very long periods of time. Other problem in higher dimensions is the 
difficulty to achieve the same resolutions used in 1D simulations due to 
the lack of computational resources, since the memory requirements in 3D 
unigrids forces one to sacrifice either the size of the physical domain or 
the numerical resolution in regions where the gravitational field still 
strong. Moreover, the number of variables in higher dimensions is also 
bigger than in 1D.\\

The practical concern is that in 3D the constraint equations are elliptic, 
which one does not want to solve at every time step due to the extreme 
hardware requirements needed to carry out a significantly long simulation. 
Therefore it is usually preferable to use unconstrained gauges and 
evaluate the constraints just to monitor the accuracy and convergence of 
the solution one is calculating. Even though unconstrained evolutions of 
Einstein's equations are still under study, it is known that some 
formulations work better than others when using certain gauge conditions. 
In this paper the BSSN formulation of general relativity is considered 
\cite{bssn}, which has shown to be more stable for longer evolutions than 
the ADM formulation for a number of physical systems \cite{towards} and 
presents different stability properties \cite{stability_ADM_vs_BSSN} (for 
test beds and comparisons under different gauges see \cite{mexico1}).\\

The Lagrangian density from which the Einstein and Klein-Gordon equations 
describing BSs are derived reads:

\begin{equation}
{\cal L} = -\frac{R}{16 \pi G} + g^{\mu \nu}\partial_{\mu} \phi^{*} 
\partial_{\nu}\phi + V(|\phi|^2),
\label{eq:lagrangian}
\end{equation}

\noindent where $R$ is the Ricci scalar, $\phi$ is a complex scalar field 
minimally coupled to gravity, $\phi^{*}$ is its complex conjugate and $V$ 
the potential of self-interaction of the field. The resulting Einstein's 
equations are $G_{\mu\nu} = 8 \pi G T_{\mu\nu}$, provided the stress 
energy tensor is

\begin{equation}
T_{\mu \nu} = \frac{1}{2}[\partial_{\mu} \phi^{*} \partial_{\nu}\phi +
\partial_{\mu} \phi \partial_{\nu}\phi^{*}] -\frac{1}{2}g_{\mu \nu}
[\phi^{*,\alpha} \phi_{,\alpha} + V(|\phi|^2)].
\label{set}
\end{equation}

\noindent
The equation satisfied by the scalar field can be obtained either from the 
Bianchi identities or by directly using the variational principle on 
(\ref{eq:lagrangian}), which for the present case reduces to the 
Klein-Gordon equation on the space-time background,

\begin{equation}
\left(
\Box - \frac{dV}{d|\phi|^2}
\right) \phi = 0
\label{kg}
\end{equation}

\noindent where $\Box \phi = \frac{1}{\sqrt{-g}}\partial_{\mu}[\sqrt{-g} 
g^{\mu \nu}\partial_{\nu} \phi]$, based on the corresponding line element:

\begin{equation}
ds^2 = -\alpha({\bf x},t)^2 dt^2 + \gamma_{i j}({\bf x},t) (dx^{i} 
+ \beta^{i} dt) (dx^{j} + \beta^{j} dt)
\label{3metric}
\end{equation}

\noindent where latin indices stand for the spatial coordinates and ${\bf x}$
indicates dependence on all spatial (cartesian) coordinates, $\alpha$ is
is the lapse function, $\beta^{i}$ the shift vector and $\gamma_{ij}$ is
the 3-metric in the 3+1 decomposition of the space-time. As I am going 
to deal only with spherically symmetric Boson Stars, a zero 
shift is used for stable equilibrium configurations, and a simple but 
sophisticated driver involving a non-zero shift will prove to be necessary 
in order to follow the evolution of unstable configurations. The general 
form of the potential to be studied here is $V(|\phi|) = \frac{1}{2}m^2 
|\phi|^2 + \frac{\lambda}{4}|\phi|^4$, which has been the most studied 
case with 1D codes \cite{seidel90,seidel98,scott2000}, among 
a variety of potentials studied \cite{mielke-sch,lee-pang92,kim-lee} 
and others including very general non polynomial potentials 
\cite{diego-f}.\\

With this notation in mind in section \ref{sec:numerical_methods} 
the numerical techniques used in the simulations shown are 
described. In section \ref{sec:Stests} physical tests are presented  
and a set of production runs from which the frequencies of the oscillation 
modes found in 1D for several configurations are reproduced with the 
present 3D code. In section \ref{sec:Utests} an example of various 
unstable BSs is shown. A trick to excite non-radial modes is presented in 
section \ref{sec:radiation} through the manipulation of the refinement of 
the grid in different directions. Section \ref{sec:restrictions} refers 
to the restrictions of the present code. Finally, some comments are 
addressed in section \ref{sec:conclusions}.\\

\section{Numerical Methods}
\label{sec:numerical_methods}
\subsection{Evolution in time}

The scalar field is described by its real and imaginary parts, such that 
$\phi({\bf x},t) = \phi_1({\bf x},t) + i \phi_2({\bf x},t)$. I 
define four new real variables based on combinations of derivatives of 
the scalar field as follows: $\pi_1 + i\pi_2 := 
\frac{{\gamma}^{1/2}}{\alpha}(\partial_t \phi - \beta^i \partial_i \phi)$, 
and $\psi_{1a} + i\psi_{2a} := \partial_a \phi$ for $a=x,y,z$, where 
$\alpha$ is still the lapse function of the space-time and $\gamma$ is the 
determinant of the three metric $\gamma_{ij}$. In terms of these new 
variables the KG equation (\ref{kg}) can be rewritten as a set of first 
order evolution equations:

\begin{eqnarray}
\partial_t \psi_1{}_{a} &=& \partial_a \left(
\frac{\alpha}{\gamma^{1/2}}\pi_1
+ \beta^i \psi_{1i} \right) \label{kg1}\\
\partial_t \pi_1    &=& \partial_i \left(\alpha \gamma^{1/2} \gamma^{ij} 
\psi_{1j} + \beta^i \pi_1\right)  -
\frac{1}{2}\alpha\gamma^{1/2}\frac{\partial
V}{\partial |\phi|^2}\phi_1 \label{kg2}\\
\partial_t \psi_2{}_{a} &=& \partial_a \left(
\frac{\alpha}{\gamma^{1/2}}\pi_2 
+ \beta^i \psi_{2i}\right) \label{kg3}\\
\partial_t \pi_2    &=& \partial_i \left(\alpha \gamma^{1/2} \gamma^{ij}
\psi_{2j} + \beta^i \pi_2\right)  -
\frac{1}{2}\alpha\gamma^{1/2}\frac{\partial
V}{\partial |\phi|^2}\phi_2 \label{kg4}
\end{eqnarray}

\noindent where the summation convention is assumed over latin 
indices. Evidently the set (\ref{kg1}-\ref{kg4}) is a system of evolution 
equations that provides the derivatives of the scalar field at each time 
slice, from which it is possible to reconstruct the scalar field itself 
using the definition of $\pi_1$ and $\pi_2$. When solving the KG 
system above, it is necessary to solve the evolution Einstein's equations 
in the BSSN formulation as described in reference \cite{bssn}. Basically 
the BSSN formulation uses the evolution variables: $\Psi = 
\ln( \det \gamma_{ij})/12$, $\tilde{\gamma}_{ij} = e^{-4\Psi}\gamma_{ij}$, 
$K = \gamma^{ij}K_{ij}$, $\tilde{A}_{ij}=e^{-4\Psi}(K_{ij}-\gamma_{ij} K 
/3)$ and $\tilde{\Gamma}^{i}=\tilde{\gamma^{jk}}\Gamma^{i}_{jk}$, 
instead of the usual ADM variables $\gamma_{ij}$ and $K_{ij}$. The 
evolution equations for the new variables are described in a number of 
references, e.g. \cite{bssn}.\\

The integration of the Klein-Gordon equation (\ref{kg1}-\ref{kg4})
and the BSSN variables is performed using the method of lines with 
second order centered differencing in space. For the time integration a 
third order Runge-Kutta (RK3) algorithm is used \cite{mol}.\\

\subsection{Gauge choice}

In the case of stable equilibrium configurations the evolution of the gauge 
was carried out using zero shift and the $1+\log$ slicing condition 
\cite{gaugeconditions}, which is a particular method according to 
which the lapse evolves as

\begin{equation}
\partial_t \alpha = -2\alpha (K-K_0)
\end{equation}

\noindent where $K$ is the trace of the extrinsic curvature and 
$K_0=K(t=0)$ ($K_0$=0 in the present case). Due to the time independence 
of the metric one knows {\it a priori} that $K_{ij} = 
-\frac{1}{2\alpha}\partial_t \gamma_{ij}=0$, i.e. that $K$ has to be zero 
for equilibrium configurations of boson stars. However this is only valid 
in the continuum limit. In practice, the discretized version of the 
equations will prevent the extrinsic curvature from being exactly zero. 
Nevertheless it was observed that the components of the extrinsic 
curvature converged to such value.\\

In order to evolve unstable equilibrium configurations the $1+\log$ 
slicing condition was also used, but it was necessary an additional 
hyperbolic gamma-driver shift condition \cite{gaugeconditions}. This type 
of conditions are necessary after a black hole has formed and prevent the 
grid stretching effect followed by an unphysical growth of the horizon.
As described in \cite{gaugeconditions}, such gamma-drivers are 
diverse, but for the spherically symmetric case I deal with here, the 
following hyperbolic condition was used:

\begin{equation}
\partial_t\beta^i = \frac{3}{4}\alpha^p B^i ~, ~~~~
\partial_t B^i = \partial_t\tilde{\Gamma}^i-\eta B^i
\label{gauges}
\end{equation}

\noindent where $p$ and $\eta$ are coefficients controlling the gauge 
speeds and certain damping added to the shift respectively.

\subsection{Exterior Boundary Conditions}

A modified version of radiative boundary conditions is applied to the BSSN 
variables and to the defined first order variables $\phi$, $\psi$ 
and $\pi$. For all these variables it is assumed that they behave as 
a constant plus an outgoing radial wave at the boundaries, plus an extra 
term that helps to control initial transient effects, that is

\begin{equation}
f({\bf r},t) = f_0 + u(r-t)/r + h(t)/r^n
\end{equation}

\noindent where $r=|{\bf r}|$, $f_0$ is one for the lapse and the diagonal 
components of the three metric and zero otherwise, $h(t)$ is a function of 
$t$ and $n$ an unknown power. The reason to choose the outgoing radial 
wave boundary condition is that it has proved to be very efficient at 
absorbing waves. The differential equation in Cartesian coordinates 
corresponding to the expression above is

\begin{equation}
\frac{x_i}{r}\partial_tf + \partial_if + \frac{vx_i}{r^2}(f-f_0) \simeq 
\frac{x_ih'(t)}{r^{n+1}}.
\end{equation}

\noindent For a given $n$, the unknown function $h'(t)$ can be evaluated 
at a point away from the boundary, and substituted in the previous 
equations to evaluate any dynamical variable at the boundary 
\cite{gaugeconditions}. It was found that when the potential of the scalar 
field has zero self-interaction ($\lambda = 0$), then switching off the 
extra term ($h=0$), that is, using standard radiative boundary conditions, 
the evolution behaves properly. Nevertheless, when the self-interaction is 
non zero, it was found that a bigger power ($n=3$ or $n=5$) prevents a pulse on 
the field variables coming from the boundaries. It is worth mentioning 
that instead of using a sponge as used in spherically symmetric evolutions 
with $\lambda \neq 0$ \cite{seidel98}, the modifications mentioned here 
sufficed to evolve the shown systems with second order accuracy for a 
number of crossing times. Finally it is important to stress that in the 
case of spherical symmetry these BCs are appropriate. More care is 
required when systems with other symmetries are studied, since this 
approximation is only valid if one considers that the front signals 
propagating outwards are spherical.\\

\subsection{Interior Boundary Conditions}

After an apparent horizon has been found for the first time, a lego sphere 
(the set of points of a cubic grid contained inside a
2-sphere) is excised from its interior. Excising is a common practice when evolving 
black hole space-times 
\cite{excision,gaugeconditions,ex-dynamic,baumgarte} and has shown to be 
an essential strategy when evolving matter fields as shown recently in 
\cite{excision-use,whisky}; the main aim of using excision is that the 
code is not forced to calculate quantities in a region close to a 
singularity, where gradients are expected to be inaccurate. In the present 
case one expects the matter field to fall into the resulting horizon in 
a finite time, thus the excision region is allowed to resize according to 
the size and shape of the apparent horizon. For this purpose the apparent 
horizon finder in \cite{jthorn-ahfinder} was used every time step. A 
buffer zone of five points between the excision region and the apparent 
horizon is kept.\\

The interior boundary is the set of points delimiting the excision region.
As apparent horizons lie inside event horizons, all the signals propagate 
towards the excision region and so do the errors. This is the reason why 
the time derivative of all the evolved variables is set to be zero at the 
the inner boundary. The implementation consists of calculating the time 
derivatives of the variables one point away from the excision region in 
the direction determined by the normal; this time derivative is then used 
to evolve the variable on the excision boundary.\\

\section{Tests on the Stable Branch}
\label{sec:Stests}

\subsection{Initial Data}

\begin{figure}[htp]
\includegraphics[width=8cm]{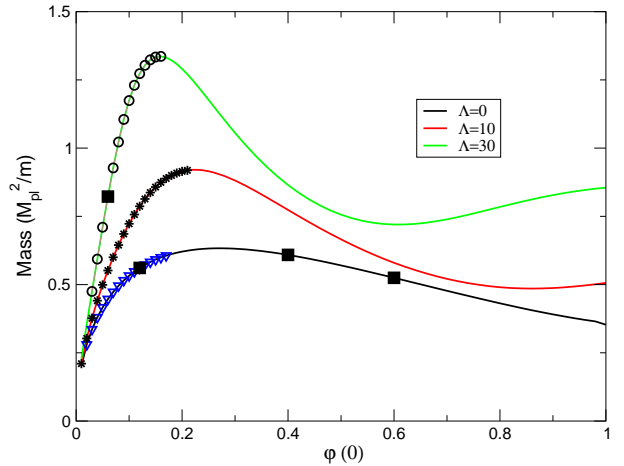}
\caption{\label{fig:eq_sequences} The sequences of equilibrium configurations
are shown for different values of the self-interaction coefficient $\Lambda$.
The configurations to the left of the respective peaks are stable 
configurations, and those to the right are unstable, which means that 
either they could collapse to form black holes or disperse away under 
infinitesimal perturbations, depending on the sign of the binding energy 
(see below). The triangles down, the stars and the circles correspond to 
the configurations evolved to trace the radial modes of oscillation of 
stable configurations. The filled squares represent four particular cases 
studied in depth to show the capabilities of the numerical techniques.}
\end{figure}

In order to test the evolution code, the first important issue is setting 
up initial data that work in 1D. For this, the well studied case of 
spherically symmetric equilibrium configurations is chosen, where the 
scalar field oscillates with constant frequency whereas the geometry in 
the chosen gauge remains 
time-independent. Einstein's equations are solved on a spherically 
symmetric background by assuming that the space time is static and that 
the scalar field has the form $\phi(r,t) = \varphi(r) e^{i \omega t}$, so 
that all the components of the stress-energy tensor are time-independent. 
Assuming $ds^2=-\alpha(r)^2dt^2 + a(r)^2dr^2 + r^2 d\Omega^2$ one arrives 
at the equations

\begin{eqnarray}
\frac{a'}{a} &=& \frac{1-a^2}{2r}+\frac{1}{4}\kappa r
\left[\omega^2 \varphi^2\frac{a^2}{\alpha^2}+\varphi'{}^{2} + a^2 V 
\right]\label{sphericalekg}\\
\frac{\alpha'}{\alpha} &=& \frac{a^2-1}{r} + \frac{a'}{a} - 
\frac{1}{2}\kappa r a^2 V\nonumber\\
\varphi '' &+& \varphi ' \left( \frac{2}{r} + \frac{\alpha '}{\alpha} - 
\frac{a'}{a}\right) + \omega^2\varphi \frac{a^2}{\alpha^2} - 
a^2 \frac{dV}{d|\varphi|^2} \varphi =0\nonumber
\end{eqnarray}

\noindent which is a set of coupled ordinary differential equations to be 
solved under the conditions $a(0)=1$, $\varphi(0)$ finite and 
$\varphi\prime(0)=0$ in order to guarantee regularity and spatial flatness 
at the origin, and $\varphi(\infty)=\varphi \prime(\infty)=0$ in order to 
ensure asymptotic flatness at infinity as described in 
\cite{ruffini,seidel90,seidel98,gleiser}. Here $\kappa = 8 \pi G$. The 
solution was calculated using the standard rescaled variables $\varphi = 
\sqrt{4\pi G} \varphi$, $r=mr$, $t=\omega t$, $\alpha = 
\frac{m}{\omega}\alpha$ and $\Lambda = \frac{\lambda}{4 \pi G m^2}$ so 
that the coordinate time is given in units of $1/\omega$ and the distance 
in units of $1/m$. In Fig. \ref{fig:eq_sequences} the sequence of 
equilibrium configurations corresponding to different self-interaction 
values is shown, and will serve to illustrate the functionality of 
this 3D code.\\

Once the system (\ref{sphericalekg}) has been solved with sufficiently 
high resolution, the straightforward identification of the 1D variables 
with the 3D variables at initial time is as follows:
$\alpha({\bf x},0)=\alpha(r)$, $\sqrt{\gamma_{rr}({\bf x},0)}=a(r)$, 
$\phi_1({\bf x},0)=\varphi(r)$, $\phi_2({\bf x},0)=0$, 
$\pi_1({\bf x},0)=0$ and $\pi_2({\bf x},0)=\frac{a}{\alpha}\varphi(r)$.
This can be easily understood just by recalling the {\it ansatz} 
$\phi(r,t)=\varphi(r)e^{it}$ in rescaled units. Then the 3D cartesian 
grid is populated by copying the values of such variables to the nearest 
neighbor grid points of the 3D domain, and applying the coordinate 
transformation from spherical to cartesian coordinates. Since it is known 
{\it a priori} that the space time should be time-independent I set the 
components of the extrinsic curvature to zero initially. Then the 
unconstrained evolution is started and the geometry and the field evolve 
according to their corresponding equations.\\

\subsection{Convergence}

\begin{figure}[htp]
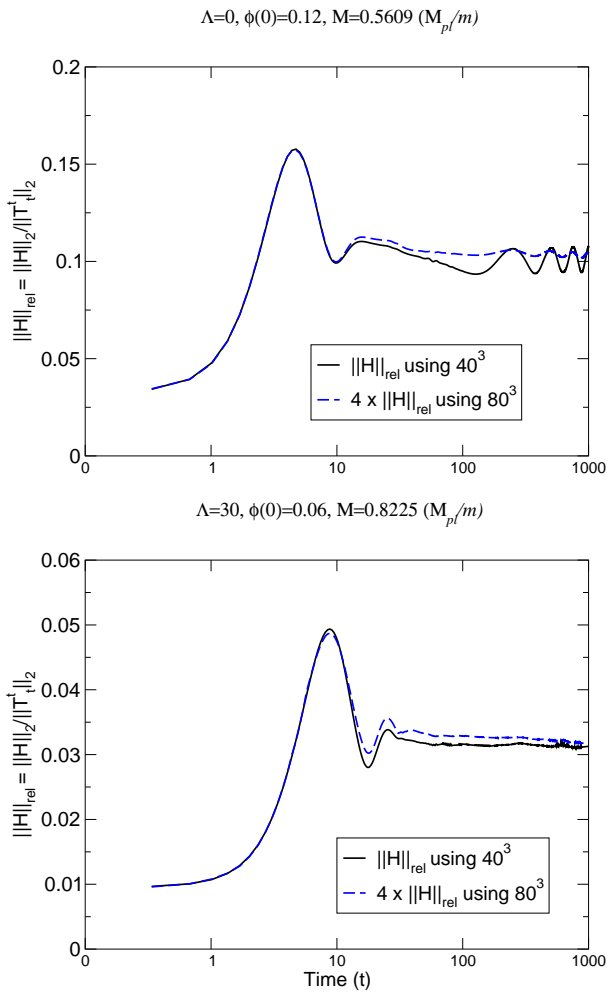

\includegraphics[width=8cm]{fig02a_hamL0.eps}
\includegraphics[width=8cm]{fig02b_hamL30.eps}
\caption{\label{fig:convergence} Convergence of the relative L2 norm of the
hamiltonian constraint for two configurations; the normalization
factor is the $T^{t}{}_{t}$ component of the stress energy tensor. 
The resolutions used were $\Delta xyx=0.675$ for the $40^3$ box 
and $\Delta xyz=0.3375$ for the $80^3$ box in each of the configurations.}
\end{figure}

The convergence of a non vacuum system is a major issue for many reasons. 
The usually applied boundary conditions to the geometrical quantities 
correspond to outgoing waves and moreover the boundary conditions for the 
scalar field with a potential term have no reported solution up to now for 
a 3D grid, so that the boundary conditions are not necessarily consistent 
with the evolution equations; these difficulties are not particular to the 
present problem, but are also a concern when solving the vacuum Einstein's 
equations or relativistic hydrodynamical fields, because these conditions 
could allow constraint violating modes to come into the domain and destroy 
the simulations. Nevertheless it was found that the modified Sommerfeld 
boundary conditions mentioned in the previous section, work well for the 
variables describing the field and the geometric quantities at once for 
quite a long time. In Fig. \ref{fig:convergence} it is shown the 
convergence of the $L_2$ norm of the violation of the Hamiltonian 
constraint produced for two initial configurations. Second order 
convergence is manifest indicating the accurate and stable behavior of the 
simulations for dozens of crossing times.\\

\begin{figure*}
\includegraphics[width=7cm]{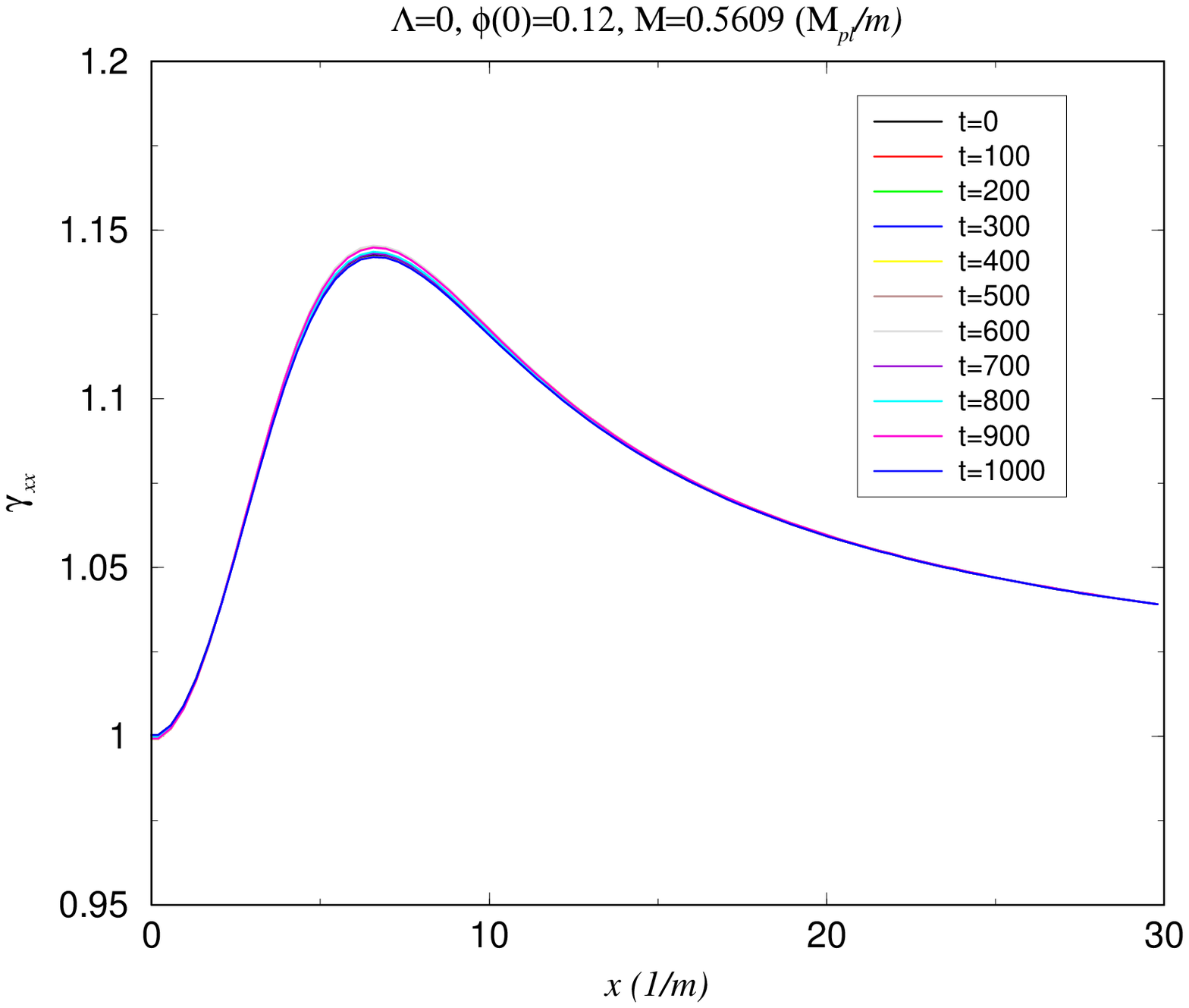}
\includegraphics[width=7cm]{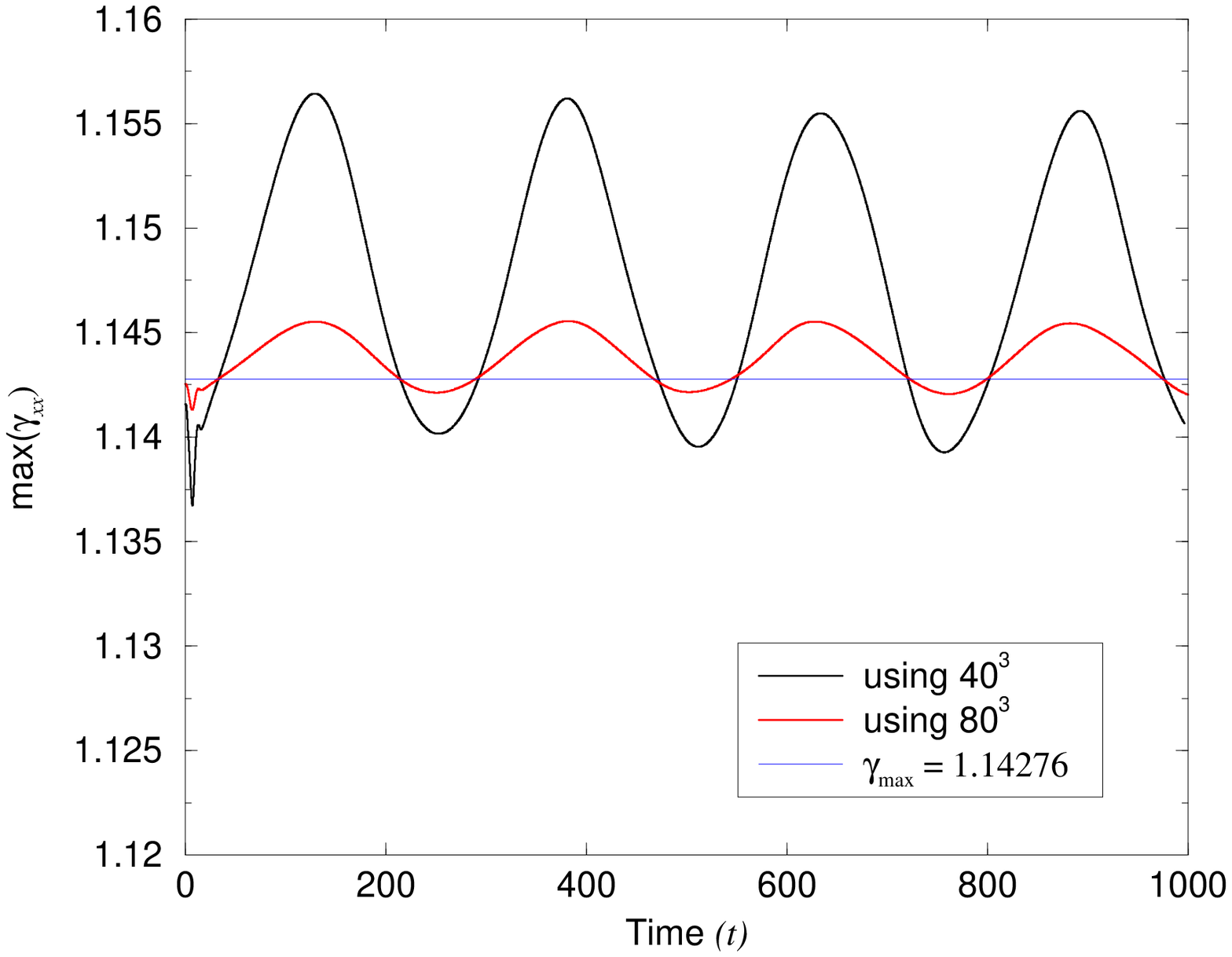}
\includegraphics[width=7cm]{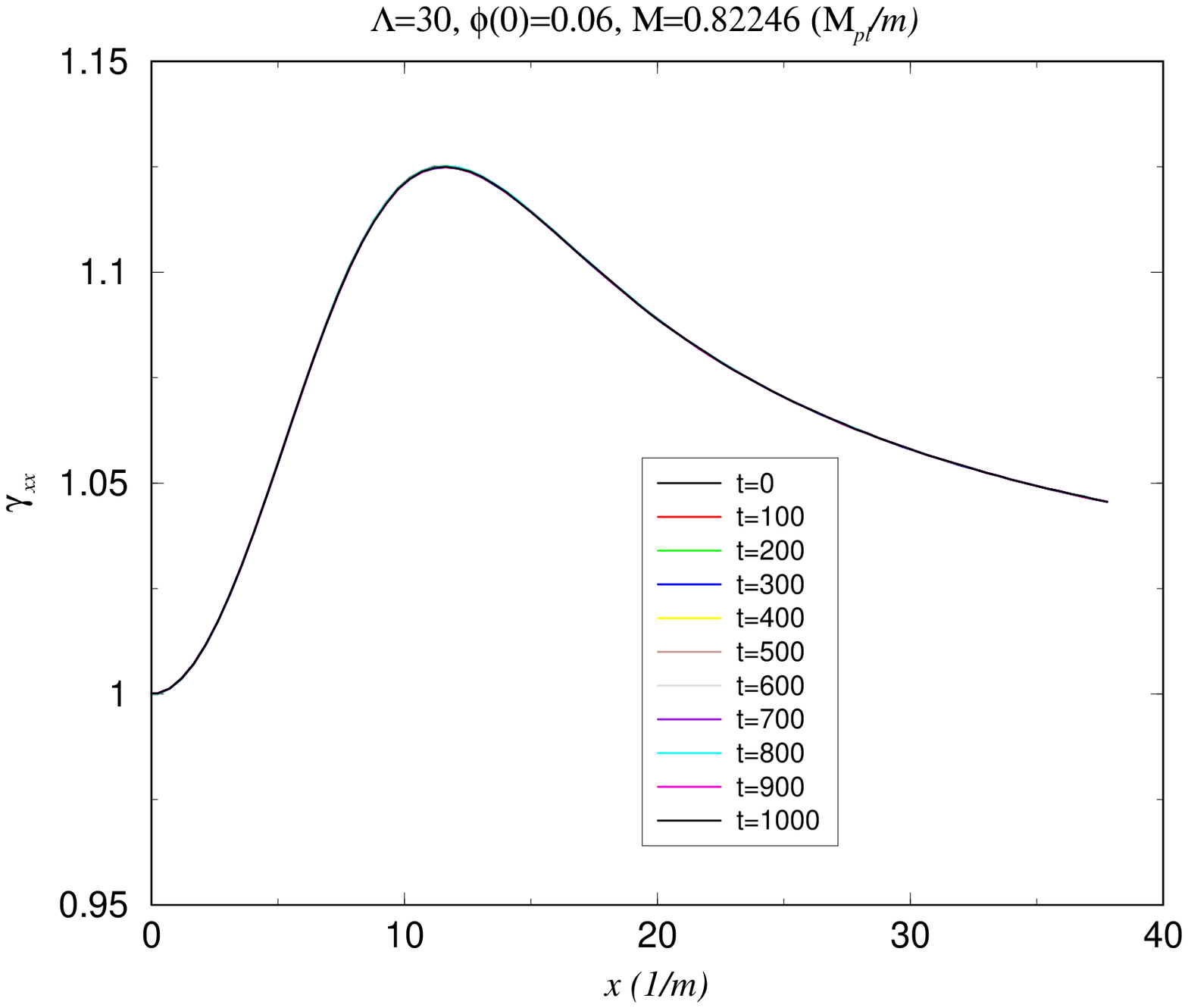}
\includegraphics[width=7cm]{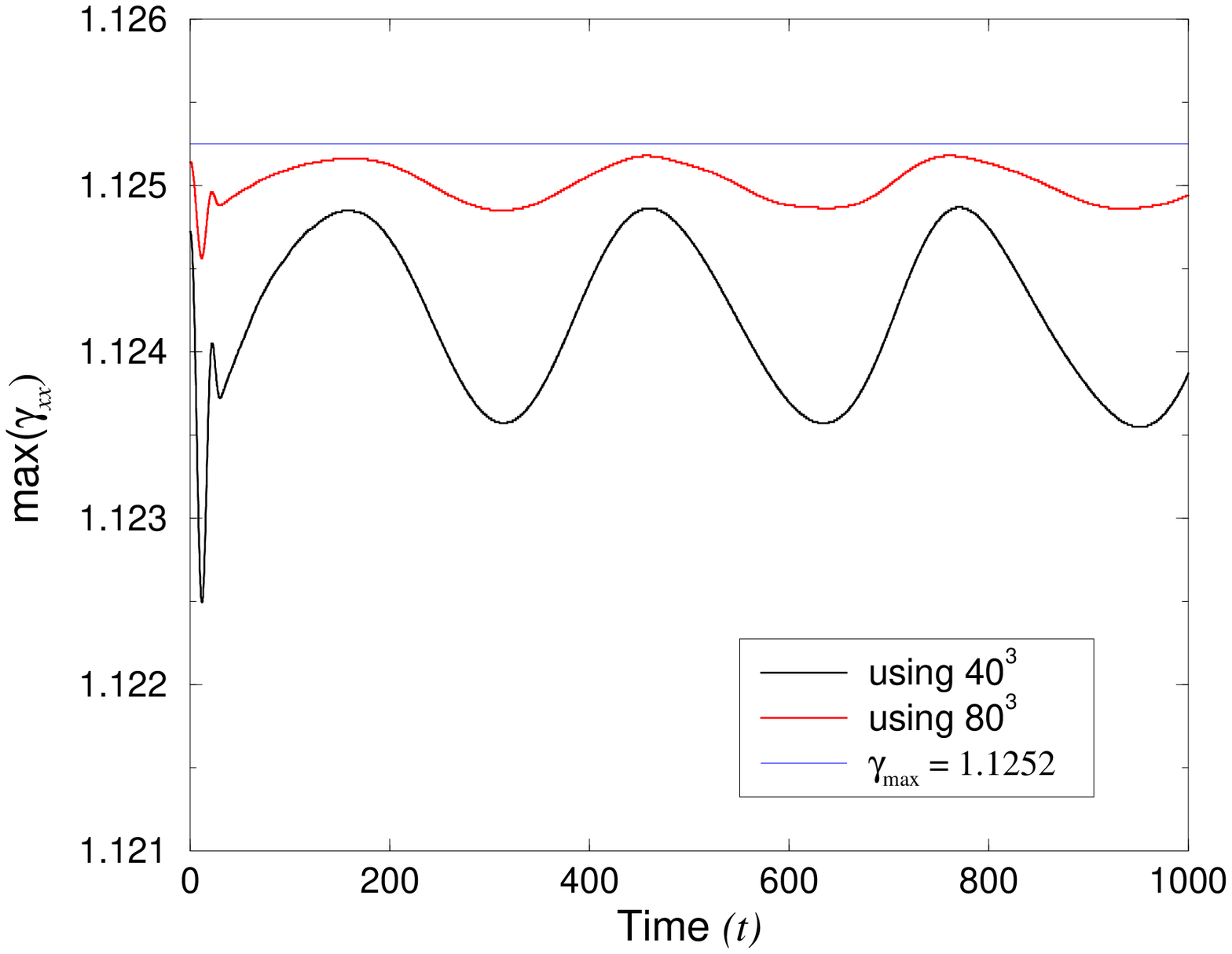}
\caption{\label{fig:t_independence} On the left, the $\gamma_{xx}$ 
metric function is shown for different systems evolved on an $80^3$ grid 
and the highest resolution mentioned in the previous figure. On the right 
is shown the shape of the maximum of $\gamma_{xx}$ in time and its second 
order convergence to the constants  $max(\gamma_{xx})=1.14276$ and 
$max(\gamma_{xx})=1.12525$, which correspond to the values for each of the 
two configurations calculated at initial time with high resolution using a 
1D code.}
\end{figure*}

One intriguing issue has to do with the time independence of the metric, 
which actually oscillates around or close to a certain fixed shape. In 
fact, due to the considerable amplitude of such oscillations it is 
possible to imagine that probably the evolved configuration corresponds to 
another configuration with different properties than those of the initial 
data. In Fig. \ref{fig:t_independence} two different cases that are 
representative are shown: in the first case, the maximum of the metric 
function $\gamma_{xx}$ oscillates around the right value calculated with 
the 1D code, which is assumed to be the correct value; in the second case 
the maximum of the metric component oscillates, but not around the 
expected value, instead with smaller values all the way. The fact is that 
in both cases the oscillating values of $max(\gamma_{xx})$ converge to 
a constant value that coincides with the one calculated at initial time. 
The second order convergence of this quantity indicates that the desired 
configuration is the one being evolved.\\

\subsection{Radial Modes}

With high spatial resolution it is quite easy to obtain very small 
variations of the metric functions in time, but since one is dealing with 
a coarse 3D grid the situation is slightly different. Instead, due to the 
errors introduced when populating the 3D grid with the initial data and 
those due to the discretization, one is dealing with perturbed 
configurations and thus it is possible to see only how the geometry 
converges to a fixed one when the spatial resolution is increased. 
Nevertheless I take advantage of this fact now to study oscillating radial 
modes as done in \cite{seidel98}, but in the present case using the 3D 
code.\\

When studying the 1D problem with very refined grids it was necessary to 
set up an explicit perturbation by adding particles or kinetic energy to 
the initial equilibrium configuration in order to study the oscillations 
of the space time \cite{seidel98} \footnote{Nevertheless this author has 
measured the same effects without the need of perturbing the number of 
particles and just taking care of the perturbations due to the 
discretization errors in a personal 1D code.}. Taking advantage of having 
a coarse grid, one gets for free the errors introduced when setting up the 
initial data, plus the discretization errors due to the numerical methods, 
act as a spherical perturbation, since an equally spaced grid is used in 
all directions, so that the evolution starts with perturbed configurations 
that converge to a strict equilibrium configuration, and it is possible to 
follow the oscillations of the metric functions. An indication that the 
perturbations applied are spherical consists in showing the amplitude of 
the Zerilli function extracted from the evolutions for the case of an 
unequally spaced grid in the $z$-direction as shown below.\\

In Fig.~\ref{fig:qqmodes} appear the results found for equilibrium 
configurations evolved for three different values of self-interaction. 
Each point represents a run on a {\bf $40^3$} (the coarse grid here) 
in octant mode of one of the equilibrium configurations indicated in 
Fig~\ref{fig:eq_sequences}, which was run up to $t\geq 1000$ in coordinate 
time and stopped because of two factors: a) even if the evolution 
algorithm used is less dissipative than the usual second order ones, 
dissipation starts being noticed and b) the phase of the scalar field also 
starts to be shifted and begins to be less good. The frequency of the 
modes shown was calculated as usual $f=1/T/\alpha(\infty_{grid})$ where 
$T$ is the period of the maximum of the metric functions and 
$\infty_{grid}$ is the outermost point of the grid. The curves shown in 
Fig.~\ref{fig:qqmodes} mimic those found for these same configurations 
obtained with a 1D code \cite{seidel98}. An important difference between 
the method used in \cite{seidel98} and the used here is that in the former 
case the system was perturbed with a shell of particles falling towards 
the star, and in the present case the discretization error is used as the 
perturbation. In the first situation the metric functions oscillate with 
an amplitude that approaches exponentially towards their values 
corresponding to the unperturbed configuration, and in the present case no 
extra particles are added, which implies that there is no extra matter to 
be ejected, and no exponential decay of the amplitude of the oscillations 
is expected, instead, the discretization error is present over all the 
numerical domain at every time.\\

\begin{figure}[htp]
\includegraphics[width=8cm]{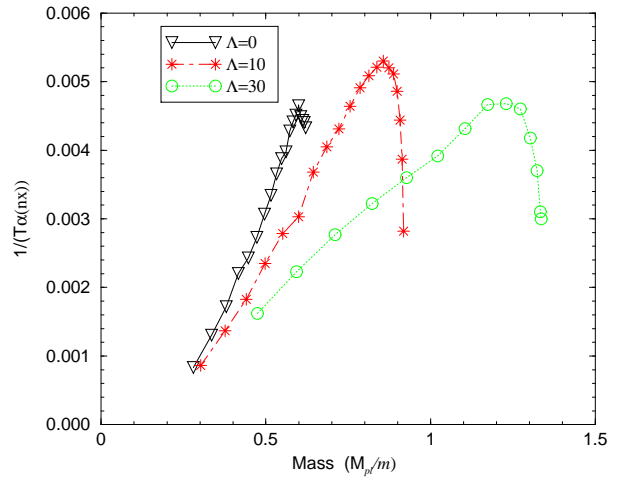}
\caption{\label{fig:qqmodes} The frequencies of oscillation of 
max($\gamma_{xx}$) for several equilibrium configurations is shown. 
It was found to be in very good agreement with those plots found with a 1D 
code for the same systems and others in reference \cite{seidel98}. Here I 
just present some cases corresponding to certain values of $\Lambda = 
\lambda / 4 \pi G m^2$.}
\end{figure}

\section{The Unstable Branch}
\label{sec:Utests}

Unstable boson star configurations are equilibrium configurations 
characterized by the fact that even using very fine tuned initial data and 
very high resolution, the finite differencing error in the calculations 
suffices to make the star collapse or disperse. In order to analyze the 
physical properties of unstable BSs I study in particular the case 
$\Lambda=0$. An important feature of the unstable branch of BSs is the 
threshold related to the sign of the binding energy $E_B = M-Nm$, where 
$N$ is the number of particles $N=\int \phi^{*} \phi$ and $M$ is the mass 
measured by an observer at infinity. If $E_B < 0$ the system is expected 
to collapse and form a black hole, and otherwise the system is unstable 
due to the excess of energy translated into kinetic energy, and (like in a 
fission process) it should be dispersed into free particles at infinity 
\cite{gleiser}, and this is precisely what is confirmed in these 
simulations\footnote{In \cite{seidel90} it was found that independently 
of the sign of the binding energy, the configurations belonging to the 
unstable branch collapse into black holes for the case $\Lambda=0$ 
studied here. In \cite{seidel98} however, it is confirmed that for $E_B>0$ 
in the case of excited boson stars the configurations are dispersed 
away. In the present 3D case, due to the limitations on the numerical 
domain, it was not possible to confirm is after an expansion (far away 
from the 3D box) the system should have recollapsed to form a black hole, 
or if a migration to a very dilute configuration ocurred. In any case, it 
is a coincidence that stars with $\phi(0) > 0.54$ simply dispersed 
away, whereas those with $0.271 < \phi(0) < 0.54$ simply collapsed without 
any noticeable ejection of matter. An essential difference to be kept in 
mind is the nature of the perturbations applied here, where the number 
of particles is not being modified at all, and the perturbation is 
applied to the system across the whole numerical domain through 
discretization error}. The limiting situation $E_B=0$ occurs for the case 
$\Lambda = 0$ when $\phi(0)=0.54$. This threshold splits the unstable 
branch of equilibrium configurations into two regions: one corresponding 
to configurations with central field value $0.271 < \phi(0) < 0.54$, i.e. 
between the critical point $\phi(0)=0.271$ separating the stable and the 
unstable branches and the threshold of zero binding energy 
$\phi(0)=0.54$; the fate of the configurations belonging to this region is 
the formation of a black hole. The second region corresponds to 
configurations with $\phi(0) > 0.54$, which should disperse away.\\

\begin{figure*}
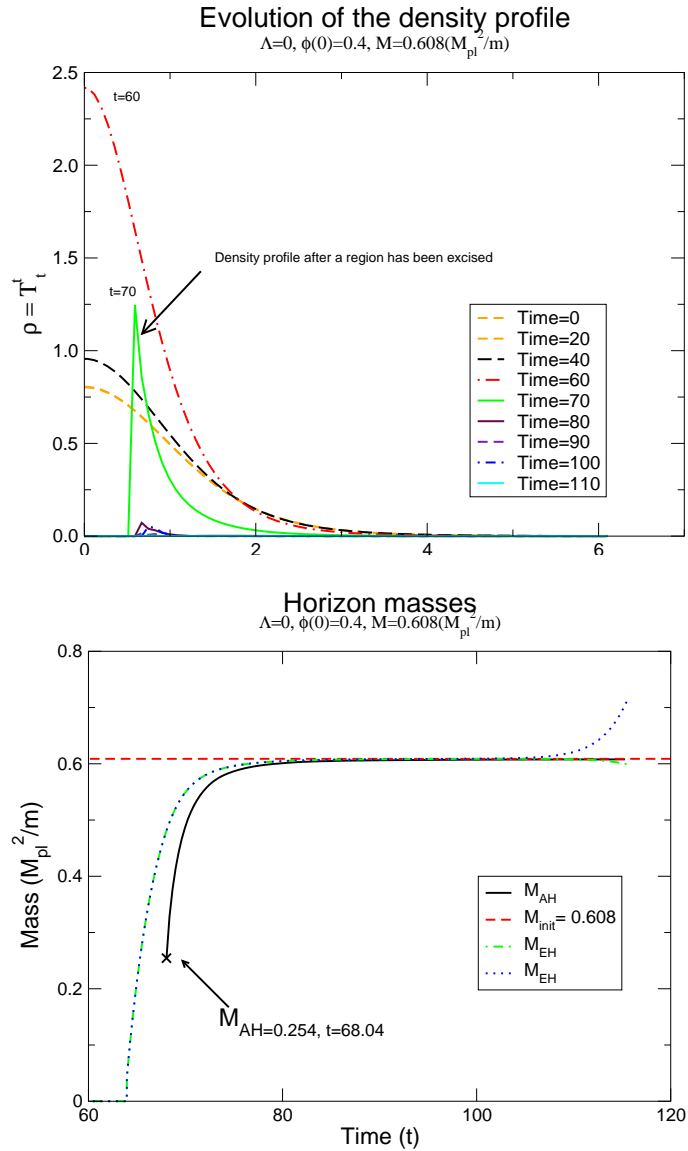

\includegraphics[width=9cm]{fig05a_density-40-new.eps}
\includegraphics[width=9cm]{fig05b_masses-40.eps}
\caption{\label{fig:star40} In the top panel the evolution of the energy 
density is shown. At the first stage of the evolution (between $t=0$ and 
20) the density remains time independent; nevertheless after $t \sim 40$ 
the distribution starts concentrating close to the origin, and after 
$t\sim 64$ the system collapses and an event horizon is formed. Then 
around $t\sim 68$ an apparent horizon is found, after which 
excision is applied. At later times the remaining matter outside the 
horizon keeps falling into it. In the bottom panel the mass of the 
apparent horizon ($M_{AH}$), the event horizon ($M_{EH}$) and the initial 
total mass ($M_{init}$) are compared. After $t=90$ and towards the end of 
the simulation the discrepancy is less than 0.4\% between $M_{EH}$ and 
$M_{init}$; a zoom in would show that the $M_{AH}$ is always less than 
$M_{EH}$. The point marked with a cross indicates the moment at which an 
apparent horizon appears, after which, the increase in its mass is due to 
the accretion of the scalar field remaining outside. The two lines 
indicating the mass of an event horizon should be understood as follows: 
an interior and an exterior initial guesses for tracking the event horizon 
are chosen, thus it is expected they converge to two different null 
surfaces at the beginning of the tracking (that is, at the end of the 
simulation); however we know that when these null surfaces coincide the 
event horizon has been hunted; therefore the event horizon starts being 
tracked at around $t=105$. The configuration corresponds to $\Lambda=0$ 
and $\phi(0)=0.4$, which is a typical case of the simulations collapsing 
into a black hole.  The gauge parameters in (\ref{gauges}) are $p=2$ and 
$\eta=0.25$. The evolution was performed with resolution $\Delta xyz = 
0.07875$ in an $80^3$ box in octant mode. The simulation was stopped at 
around $t \sim 117$ due to an instability coming from the inner boundary.} 
\end{figure*}

In Fig. \ref{fig:star40} a typical density evolution is presented 
for the case of a collapsing BS, both before and after the Black 
Hole is formed. After finding an apparent horizon for the first time, the 
excision is applied to all the variables that are being evolved and those 
serving to monitor the system (e.g. the energy-density). In the same 
figure the mass of the apparent and event horizons of the resulting black 
hole are shown. The discrepancy between the mass of the apparent and event 
horizons and the original mass of the system is within 0.4\% after most of 
the matter has been absorbed, around $t \sim 90$. The event horizon is a 
concept that depends on the global structure of the space-time, therefore 
it can only be tracked in a post-process step after the simulation was 
finished. The event horizon finder \cite{ehfinder} works by evolving null 
surfaces backwards in time using the information stored during the 
space-time evolution. Two initial surface guesses are necessary to start 
the event horizon tracking. One guess is chosen in a region outside and 
other inside the expected location of the horizon. These guesses were 
chosen based on the apparent horizon radius. The event horizon is at the 
end the surface the two initial guesses converge to. This is the reason 
why in Fig. \ref{fig:star40} two different event horizon masses are shown. 
It can be seen that by the end of the simulation (the start of the event 
horizon tracking) the mass of the two initial surfaces does not coincide, 
and that around $t=105$ the two measurements converge to a single 
one: the event horizon.\\

\begin{figure}[htp]
\includegraphics[width=8cm]{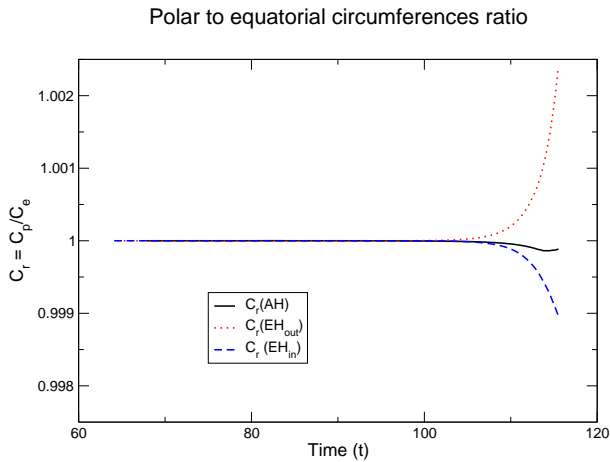}
\caption{\label{fig:circ} The polar to equatorial circumference ratio is 
shown for the different horizons, and it is manifest that their shape is 
spherical within an error of 0.2\% towards the end of the simulation.} 
\end{figure}

\begin{figure}[htp]
\includegraphics[width=8cm]{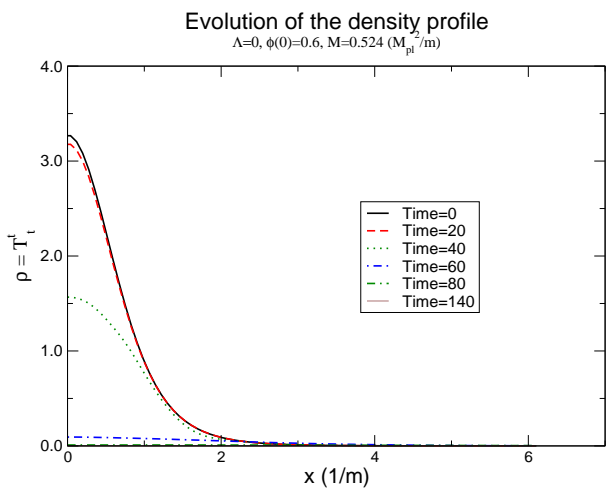}
\caption{\label{fig:dispersed} 
The energy density profile of the configuration $\Lambda=0$ and 
$\phi(0)=0.6$ is shown at different stages. Due to the restricted domain 
it could happen either that the system should recollapse after traveling 
a distance outside the box to form a black hole as claimed in 
\cite{seidel90}, or it is confirmed the fact that for configurations with 
positive binding energy the system truly disperses away and never 
recollapses. This controversy is expected to be solved once the present 
code is able to use notably bigger domains with the help of mesh 
refinement techniques. The configuration corresponds to $\Lambda=0$ and 
$\phi(0)=0.6$. The gauge parameters and resolution are the same as those 
used for the collapsing system shown above.} 
\end{figure}

Based on these two important tools it is shown in Fig.~\ref{fig:circ} that 
the ratio of polar to equatorial circumferences is one over the evolution, 
indicating that the horizons are spherical, despite the lego-type shape of the 
excision boundary. This ratio will be fundamental in determining the 
physical parameters of the resulting black hole in the case of evolutions 
with fewer symmetries. E.g. from the oscillations of $C_r=C_p/C_e$ it is 
possible to obtain the angular momentum of a black hole resulting from the 
collapse of very general configurations (see e.g. 
\cite{brandt-seidel,coalescence}). Finally, in Fig. \ref{fig:dispersed} 
the evolution of $T^{t}_{t}$ for a dispersing BS is presented, where 
the ejection of matter is manifest.\\

\section{Exciting non-radial modes}
\label{sec:radiation}

\begin{figure}[htp]
\includegraphics[width=8cm]{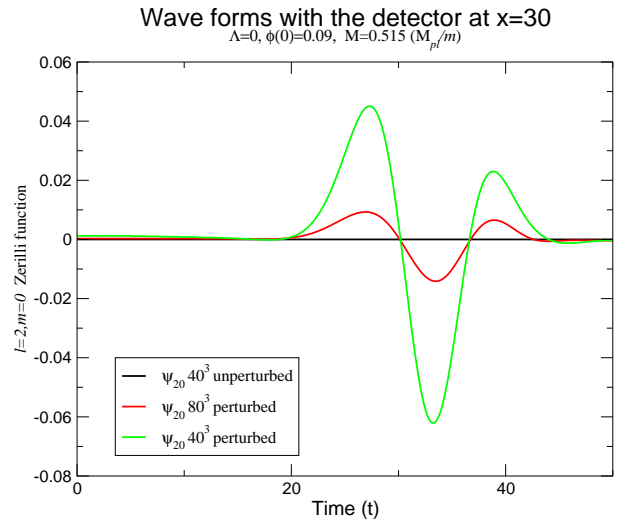}
\caption{\label{fig:radiation} The wave form and its convergence
is shown for the fundamental mode with a detector located at
$x=30(1/m)$. The perturbed system was $\Lambda=0$, $\phi_1(0,0)=0.09$ and 
$M=0.51533 (M_{pl}/m)$. The solid line around zero is the amplitude
of the wave form when the system is evolved on the isotropic grid.}
\end{figure}

Since the systems presented here have spherical symmetry, no gravitational 
wave signals are expected. Nevertheless as an exercise and in order to 
show that the modes studied in section~\ref{sec:Stests} were actually 
radial, I take advantage of the freedom one has in 3D to apply a simple 
trick just by choosing different spatial resolutions $\Delta x=\Delta y 
\ne \Delta z$ with the consequent effect that the system is being 
perturbed in the $z$ direction, and thus gravitational radiation is 
obtained in the fundamental mode simply due to discretization error.
The perturbation consisted in using a grid size $40 \times 40 \times 35$ 
in the $x,y,z$ directions respectively and resolution such that $\Delta x 
= \Delta y$, $\Delta z$ was chosen so that the range in the three 
directions was the same. In Fig.~\ref{fig:radiation} the waveform for the 
fundamental mode $l=2$, $m=0$ are shown and compared with the results 
found for an isotropic grid. The wave forms were extracted with the 
methods described in \cite{gab}. The convergence of the wave forms is 
second order, and as expected it converges to zero, since in the continuum 
limit it is expected the grid to be isotropic and no signal in this mode 
could come out from a spherically symmetric source. In this way it is also 
shown that the modes analyzed in the section \ref{sec:Stests} were 
actually radial.\\

\section{Features of the code}
\label{sec:restrictions}

\subsection{Restrictions}

The period of time that the convergence is preserved acts as an indicator
of the consistency properties of the scheme, formulation, and boundary 
conditions working all together. The runs were stopped at the times shown 
in the plots because: a) the convergence was not manifestly second order 
with the passage of time, b) the period of the scalar field drifted after 
a finite time. These problems could be sorted out just by using higher 
resolution in time; all the simulations were carried out using a CFL 
factor of $\Delta t/\Delta x = 0.05$ (which means that the runs required 
tens of thousands of iterations), but convergence and period could be 
preserved just by decreasing such number. Another reason to choose such 
a small value of the CFL factor was that it helps at diminishing 
dissipation after a very long time of evolution.\\

Another issue is the problem of resolution. During the evolutions 
presented in this manuscript there were at most a couple of dozens of 
points (using the low resolution) covering the volume containing 95\% of 
the total mass of the star, and twice as much outside such region. In 
consequence the gravitational wave detectors had to be placed rather close 
to the BS. But even so, the results are good. The ideal situation for more 
general configurations would be: high resolution in the central part of 
the grid and a physical domain big enough to allow one to place detectors 
far away from the source. Thus practically means that the detector would 
be located in a Schwarzschild-like region thus providing cleaner 
signals.\\

Using a uniform grid works fine for spherical symmetry, but presents 
a dilemma in other cases: use high resolution to cover the central part 
and then forget about obtaining high quality waveforms, or conversely, 
sacrifice the resolution at the center and pull the grid far away in order 
to have a clean gravitational wave signal, but produced by a not 
accurately evolved source. The solution to this problem would be provided 
by the implementation of mesh refinement \cite{carpet,nasa,liebeling}, 
and will be useful when dealing with systems under more interesting 
symmetries.\\

\subsection{Restrictions on the physics}

The phase of the field is an important indicator of the control one has on 
the simulation, because if the field is out of phase or the period of 
oscillation of the field are not the right ones, then the effects will be 
noticed in the geometric quantities. Among the experiments carried out, 
it was observed that when using a three step iterative Crank-Nicholson 
(3ICN) \cite{teukolsky00a} or second order Runge-Kutta (RK2) methods to 
evolve the system, the period of the scalar field changed after a while 
($t \sim 200$), whereas the third order Runge-Kutta preserved the phase 
and the period, although a modified version of ICN \cite{oscillatons} 
proved to work properly as well.\\

The dissipation is another issue one has to deal with when evolving
matter. By observing the amplitude of the scalar field it was found that 
after $t \sim 1000$ the 3ICN provided an amplitude of the field $\sim 
10$\% less than the original one, while with the RK2 the amplitude of the 
field it was $ 15\%$ bigger than the right value for the resolutions used 
here, which effects were evident in the measurement of physical 
quantities. The amplitude of the field after $t \sim 1000$ was less than 
the original by 0.3\% using the coarse resolution. Thus, the RK3 was the one providing the 
most 
confident results in terms of convergence for longer time, dissipation, 
phase and period of oscillation of the solutions. It was
also essayed the fourth order Runge-Kutta scheme, however it 
required more time iterations and no major improvement in accuracy was
found. RK3 showed to be optimal in performance and accuracy for the
purposes of the runs shown here.\\

In the case of stable configurations, the duration time scale of the 
simulations was enough to observe the oscillations of a wide range 
of stable configurations. However, the expected period of oscillation 
for very diluted configurations (those to the left in Fig. 
\ref{fig:qqmodes}) is bigger if the mass is smaller. Therefore, it was not 
possible to study those configurations within the ranges of evolution 
times achieved with the present resolution and evolution times at the 
moment.\\

In the case of unstable configurations it is not expected to observe a 
constant amplitude of the scalar field, instead the squeezing of the
scalar field, which is confined to a small region, implies an increase in the
amplitude of the field and energy density whereas it is more concentrated
in a small region where nodes can be observed before the formation of an 
apparent horizon. After excision is applied, the gradients of the scalar 
field become important in the region close to the inner boundary, and 
some unstable modes develop at some point. We are certain these modes 
are not numerical artifacts because increasing the resolution they last 
more time to show up. I expect to report about this point in further 
reports.\\

\section{Comments and Conclusions}
\label{sec:conclusions}

I have presented a 3D code that is able to keep under control the 
unconstrained evolution of spherically symmetric Boson Stars, both stable 
and unstable using the BSSN formulation of General Relativity, the 
$1+\log$ slicing condition, a coarse grid and simple discretizations on a 
uniform grid and well known numerical methods to evolve in time.\\

Stable configurations were evolved during several crossing times with 
second order convergence of the hamiltonian constraint and some 
characteristic properties of the geometry, like the convergence of the 
maximum of the radial metric function to the appropriate value. The 
process of collapse of unstable configurations was followed for quite a 
long time of around $t \sim 80 M$, after the BH was formed, with error in 
the mass of the horizons of less than 0.4\% in several cases studied. The 
configurations that dispersed where followed for a very long time and no 
surprises were found when evolving the resulting flat space-time.\\

It has been shown that these evolution techniques work properly when 
the IVP is not solved in the grid used for the evolution. This procedure 
will permit one to distinguish future problems intrinsically related to 
the solution of the initial value problem directly on the 3D grid 
\cite{ruxandra}. I also decided to use simple uniform grids in order 
to set up test beds for future Boson Star simulations without involving 
coordinate transformations \cite{fisheye} that will be used for further 
research or mesh refinement that will be helpful as well \cite{carpet}.\\

It is possible to conclude by stating that it has been possible to 
reproduce in a 3D cartesian grid, the basic results found with 1D codes 
for spherically symmetric Boson Stars, which are related to infinitesimal 
perturbations due to the discretization error. In addition, taking  
advantage of the freedom one has in 3D grids to choose the discretization 
error in different directions, it was possible to extract a wave form that 
illustrates the type of signals one could expect from BSs and showed that 
the signal obtained when exciting the fundamental mode, presents a quick 
damping as predicted for axial perturbations of Boson Stars 
\cite{yoshida-erigichi-futamase}.\\

A comment about the late development of a 3D code for boson stars is in 
turn. Under the ADM formulation of General Relativity \cite{adm} it was 
rather difficult to evolve a Boson Star in 3D for the periods of time 
shown here (see e.g. \cite{towards}). The main reason preventing such long 
evolution was the ADM formulation itself, which showed to be inefficient 
at evolving a garden variety of physical systems \cite{towards}. On 
the other hand, at the present time one accounts with several new gauge 
conditions allowing the control of the simulations. A few years ago 
considerable effort was invested to evolve Boson Stars in 3D that did not 
succeed only because of the formulation used \cite{choi,jay}, although 
previous studies using new formulations were also explored \cite{thomas}. 
It is necessary to point out that now it is possible to study in depth the 
phenomenology of Boson Stars in full 3D with the new techniques and tools 
that are emerging in the field of numerical relativity. Boson Stars could 
still be systems helpful at testing new numerical techniques, but most 
important they can be considered objects whose existence could eventually 
be probed if they are thought to be sources of gravitational waves.\\

This code was written as a thorn in the Cactus frame, and can be plugged 
into the variety of analysis tools and drivers available in the Cactus 
toolkit \cite{cactus-tk}.\\


\begin{acknowledgments}
The author thanks Horst Beyer, Peter Diener, Ian Hawke, Ed Seidel, Arturo 
Ure\~na and Jason Ventrella for important comments and suggestions;  
special thanks to Ian Hawke for implementing the MoL thorn, Jonathan 
Thornburg for implementing the apparent horizon finder, Peter Diener 
for implementing the event horizon finder, and the Cactus-team for 
providing the necessary development tools. The Cactus and CactusEinstein
infrastructure with AEI thorns were extensively used. The simulations were 
carried out in the Peyote cluster at the AEI and in the SuperMike 
cluster at LSU. This work was supported by the Center for Computation and 
Technology at LSU. This research was partly supported by the 
German-Mexican bilateral project DFG-CONACyT 444 MEX-13/17/0-1.
\end{acknowledgments}





\begin{thebibliography}{}

\bibitem{ruffini} R. Ruffini and S. Bonazolla, Phys. Rev. {\bf 187} 
        (1969) 1767.
\bibitem{kaup} D. J. Kaup, Phys. Rev. {\bf 172} (1968) 1331.
\bibitem{colpi} M. Colpi, S. L. Shapiro and I. Wasserman, Phys. Rev. Lett.
        {\bf 57} (1986) 2485.
\bibitem{dark-matter} Jae-weon Lee, In-guy Koh, Phys. Rev. {\bf D 53} (1996) 
        2236. E. W. Mielke and F. E. Schunck, Phys. Rev. {\bf D 66} (2002) 
        023503. F. S. Guzm\'an and L. A. Ure\~na-L\'opez, Phys. Rev. {\bf 
        D 68} (2003) 024023.
\bibitem{diego}  D. F. Torres, S. Capozziello, G. Lambiase Phys. Rev.
        {\bf D 62} (2000) 104012.
\bibitem{ryan} F. D. Ryan, Phys. Rev. {\bf D 55} (1997) 6081.
\bibitem{towards}  M. Alcubierre, B. Bruegmann, T. Dramlitsch, J. A. Font, 
        P. Papadopoulos, E. Seidel, N. Stergioulas and R. Takahashi, Phys. 
        Rev. {\bf D 62} (2000) 044034.
\bibitem{jetzer} P. Jetzer, Phys. Rep. {\bf 220} 4 (1992) 163.
\bibitem{cqgreview} F. E. Schunck and E. W. Mielke, Class. Quantum Grav. 
        {\bf 20} (2003) R301.
\bibitem{seidel90} E. Seidel and W-M. Suen, Phys. Rev. D {\bf 42}, 384 (1990);
\bibitem{seidel98} J. Balakrishna, E. Seidel, and W-M. Suen, Phys. 
        Rev. D {\bf 58}, 104004 (1998).
\bibitem{scott2000} S. H. Hawley and M. W. Choptuik, Phys. Rev. {\bf D 62} 
        (2000) 104024.
\bibitem{bssn} M. Shibata and T. Nakamura, Phys. Rev. {\bf D 52} (1995) 
        5428. T. W. Baumgarte and S. L. Shapiro, Phys. Rev. {\bf D 59} 
        (1999) 024007.
\bibitem{stability_ADM_vs_BSSN} M. Alcubierre, G. Allen, B. Br\"ugmann, 
        E. Seidel and W-M Suen, Phys. Rev. {\bf D 62} (2000) 124011.
\bibitem{mexico1} M. Alcubierre {\it et al.}, Class. Quantum Grav., {\bf 21} 
        (2004) 589.
\bibitem{mielke-sch} E. W. Mielke and R. Scherzer, Phys. Rev. {\bf D 24} 
        (1981) 2111.
\bibitem{lee-pang92} T. D. Lee and Y. Pang, Phys. Rep. {\bf 221} (1992) 
        251.
\bibitem{kim-lee} J-W Ho, S-J Kim and B-H Lee, gr-qc/9902040.
\bibitem{diego-f} F. E. Schunck and D. F. Torres, Int. J. Mod. Phys. {\bf 
        D 9} (2000) 601.
\bibitem{mol} R. J. LeVeque, {\it Numerical Methods for Conservation 
        Laws}. Series, Birkh\"auser Verlag, Basel, 1990.
\bibitem{excision} M. Alcubierre and B. Bruegmann, Phys. Rev. {\bf D 63} 
        (2001) 104006.
\bibitem{gaugeconditions}  M. Alcubierre, B. Bruegmann, P. Diener, M. 
        Koppitz, D. Pollney, E. Seidel and R. Takahashi, Phys. Rev. 
        {\bf D 67} (2003) 084023.
\bibitem{ex-dynamic} M. Alcubierre, B. Bruegmann, D. Pollney, E. Seidel 
        and R. Takahashi, Phys. Rev. {\bf D 64} (2001) 061501.
\bibitem{baumgarte} H.J. Yo, T. W. Baumgarte and S. L. Shapiro,
        Phys. Rev. {\bf D 64} (2001) 124022.
\bibitem{excision-use} M. D. Duez, S. L. Shapiro, W-J Yo, gr-qc/0401076. 
\bibitem{whisky} L. Baiotti, I. Hawke, P .J. Montero, F. Loeffler, L. 
        Rezzolla, N. Stergioulas, J. A Font, E. Seidel, gr-qc/0403029.
\bibitem{jthorn-ahfinder} J. Thornburg, Class. Quantum Grav. {\bf 21} 2 
        (2004) 743.
\bibitem{gleiser} M. Gleiser, Phys. Rev. {\bf D 38} (1988) 2376.
\bibitem{ehfinder} P. Diener, Class. Quantum Grav. {\bf 20} (2003) 4901.
\bibitem{brandt-seidel} S. Brandt and E. Seidel, Phys. Rev. {\bf D 52} 
        (1995) 856.
\bibitem{coalescence} M. Alcubierre, B. Bruegmann, P. Diener, F. S. 
        Guzm\'an, I. Hawke, F. Herrmann, M. Koppitz, D. Pollney, E. Seidel 
        and J. Thornburg, in preparation. 
\bibitem{gab} G. Allen, K. Camarda and E. Seidel, gr-qc/9806014 and 
        gr-qc/9806036. 
\bibitem{carpet} E. Schnetter, S. H. Hawley and I. Hawke, Class. Quantum 
        Grav. {\bf 21} (2004) 1465.
\bibitem{nasa} B. Imbiriba {\it et al.}, gr-qc/0403048.
\bibitem{liebeling} S. L. Liebling, gr-qc/0403076.
\bibitem{teukolsky00a} S. Teukolsky, {\it Phys. Rev. D} {\bf 61}, 087501 
        (2000).
\bibitem{oscillatons} M. Alcubierre, R. Becerril, F. S. Guzm\'an, T. Matos, 
        D. N\'u\~nez and L. A. Ure\~na-L\'opez, Class. Quantum Grav., 
        {\bf 20} (2003) 2883.
\bibitem{ruxandra} J. Balakrishna, R. Bondarescu, G. Daues, F. S. Guzm\'an 
        and E. Seidel, in preparation.
\bibitem{fisheye} J. Baker, M. Campanelli and C. Lousto, Phys. Rev. 
        {\bf D 65} (2002) 044001.
\bibitem{yoshida-erigichi-futamase} S. Yoshida, Y. Eriguchi and T. Futamase,
        Phys. Rev. {\bf D 50} (1994) 6235.
\bibitem{adm} J. York, {\it Sources of Gravitational Radiation}, Ed by Larry
        L. Smarr. Cambridge University Press, Cambridge, 1979.
\bibitem{choi} D. I. Choi, Ph D Thesis, The University of Texas at Austin 1998.
\bibitem{jay} J. Balakrishna, Ph D Thesis, Washington University 1999.
        E-version: gr-qc/9906110.
\bibitem{thomas} T. Dramlitsch, Masters Thesis, Universit\"at Potsdam, 1999.
\bibitem{cactus-tk} B. Bruegmann, Ann. Phys. (Leipzig) {\bf 9}, 3-5 
        (2000) 227. E. Seidel and W-M Suen, JCAM {\bf 109} (1999) 493. 
        {\bf http://www.cactuscode.org} 
\end{thebibliography}
\end{document}